# Effect of the transfer reactions for $^{16}$O+$^{10}$B elastic scattering

N. Burtebayev[a], Sh. Hamada[b], Awad A. Ibraheem[c,d], K. Rusek[e], M. Wolinska-Cichocka[e], J. Burtebayeva[a], N. Amangeldi[f], Maulen Nassurlla[a,g], Marzhan Nassurlla[a] and A. Sabidolda[a]

[a]Institute of Nuclear Physics, Almaty, Kazakhstan
[b]Faculty of Science, Tanta University, Tanta, Egypt
[c]Physics Department, King Khalid University, Abha, Saudi Arabia
[d]Physics Department, Al-Azhar University, Assiut Branch, Assiut 71524, Egypt
[e]Heavy Ion Laboratory, University of Warsaw, Pasteura 5A, 02-093 Warsaw, Poland
[f]L. N. Gumilyov Eurasian National University, Astana, Kazakhstan
[g]Al Farabi Kazakh National University, Almaty, Kazakhstan

## Abstract

In this study, the angular distribution of the $^{16}$O+$^{10}$B elastic scattering was measured at $E_{lab}$ ($^{16}$O) = 24 MeV. In addition to our experimental data, this nuclear system was theoretically analyzed at different energies to study the dynamics of scattering for this system. The data were analyzed within the framework of the double-folding optical potential model. The values of the spectroscopic factors (*SA*) for the configuration $^{16}$O→$^{10}$B+$^{6}$Li were extracted at the energies at which the effect of the $^{6}$Li cluster transfer on the cross-sections at backward angles is observed. The energy dependence of the reaction cross-section for this system was also investigated.

## I. Introduction

The effect of transfer of a nucleon or a group of nucleons on the differential cross sections at backward angles is well understood. The transfer phenomenon, observed in many systems, could mainly be classified two types: (*a*) nuclear systems where the entrance and exit channels are physically indistinguishable, i.e., there is only a rearrangement of the reaction products, for example, $^{12}$C ($^{16}$O,$^{12}$C)$^{16}$O "elastic transfer" and (*b*) nuclear systems where the entrance and the exit channels are different, for example, $^{6}$Li ($^{3}$He,d)$^{7}$Be "transfer reaction". The two aforementioned types of transfer processes have been extensively studied and have been found to show a significant increase in their differential cross sections at backward angles. These transfer processes could not be investigated within the framework of the optical model and consequently, the Distorted Wave Born Approximation (DWBA) method or Coupled Reaction Channel (CRC) method was found to be necessary to define these processes [1–6]. One of the quick observations that could be drawn about the interaction potentials for them is that in nuclear systems of type (*a*), the interaction potentials for the entrance and exit channel are usually the same, while for nuclear systems of type (*b*), the interaction potentials are different. Of course, other potentials should be included for overlapping and coupling, in addition to the spectroscopic amplitude. Spectroscopic factor ($C^2S$), which is the square of the spectroscopic amplitude (*SA*), is related to the preformation probability of a cluster configuration in a nucleus. Thus, extracting reliable values for $C^2S$ enables obtaining better knowledge about the nuclear structure of the interacting nuclei and the reaction mechanism. In the present work, we have measured the angular distribution of $^{16}$O elastically scattered from $^{10}$B targets at $E_{\text{lab}}$ ($^{16}$O) = 24 MeV. In addition to our experimental data, we have carried out a theoretical analysis of the data for this system at different energies in order to address its features and peculiarities. There are several experimental measurements for the $^{16}$O+$^{10}$B system. In an earlier report [7], the elastic scattering angular distributions for the $^{16}$O+$^{10}$B were measured at energies $E_{\text{lab}}$ ($^{16}$O) = 18.2, 21.37, 23.27, 26.0 and 27.3 MeV. The data in this energy range extended up to an angle of 120$^\text{o}$ and did not show any remarkable variation in the cross-sections at backward angles. These data were analyzed using the phenomenological Wood-Saxon potential. On the other hand, the data at higher energies [8–10], i.e., at energies above the Coulomb barrier, did show enhanced cross-sections at backward angles. Different explanations have been proposed for this observation, such as *a*) the contribution of the elastic transfer process, *b*) coupling to important reactions channels, and *c*) compound elastic processes. In an extensive study of the systems $^{16,17,18}$O+$^{10,11}$B and $^{19}$F+$^{9}$Be [9], the enhanced cross-sections at backward angles were observed and explained to be due to the contribution of compound nucleus processes. To the best of our knowledge the enhancement of cross sections at backward angles haven't been investigated in terms of the $^{6}$Li cluster transfer. Koide *et al.*. [8] studied the $^{16}$O+$^{10}$B elastic scattering at energies of $E_{\text{lab}}$ ($^{16}$O) = 36.58, 41.99, and 48.49 MeV and their results exhibited a significant increase in the differential cross sections at backward angles. The forward angle part of the angular distributions was described by the scattering matrix, $S_0$, and the backward angle part by an anomalous matrix, $\tilde{S}$ .They also claimed that the inclusion of the $^{6}$Li cluster transfer does not reveal a good agreement with the backward angle scattering data. In another work [10], experimental measurements for

$^{16}$O+$^{10}$B elastic scattering at $E_{lab}$ ($^{10}$B) = 100 MeV did not extend to the sufficiently backward angles to enable investigating the role of cluster transfer on the cross-sections at these angles. The current work is devoted to studying the dynamics of $^{16}$O+$^{10}$B elastic scattering as well as to extract reliable values of the spectroscopic factor for the configuration $^{16}$O→$^{10}$B+$^{6}$Li.

## II. Experimental details

The experiment was performed in the cyclotron DC–60 INP NNC Republic of Kazakhstan. The cyclotron DC–60 can accelerate ions from $^{6}$Li to $^{132}$Xe in the energy range of 0.35 MeV/n to 1.75 MeV/n The frequency of the accelerated ions is 4.22÷1.84 MHz. The range of the mass to charge ratio (A/Z) for the accelerated ions is *6–12*. The voltage applied to the Dees is 50 kV. The variation of the ion energy in the range of 0.35 to 1.75 MeV/n is ensured by changing the charge of the accelerated particles and magnetic field of the cyclotron. The power supply system of the cyclotron magnet consists of a main winding and a system of correcting coils. Mean magnetic field is in the range of 1.25 to 1.65 T, the magnet poles are of diameter 1.6m. The experiment was conducted using the scattering chamber shown in Fig. 1. The chamber is made from a single block of stainless steel with an inner diameter 430 mm and an internal height 200 mm. The vacuum system used is a turbo-molecular pump of the capacity of 250 L/s and a backing pump with a capacity of 190 L/min. The pumping system was tested and showed good results in both pumping speed and in achieving a high vacuum.

The $^{16}$O ion beam was accelerated up to the energy of 24 MeV and then directed onto a 35.3 μg/cm$^2$ foil of natural Boron target ($^{10}$B–60%, $^{12}$C–25%, $^{16}$O–15%). The dead time was monitored and kept as constant as possible by changing the spectrometer entrance slits and/or the beam intensity. Energy spectra of the elastically scattered $^{16}$O particles were measured using a silicon surface barrier detector (ORTEC) with a sensitive layer thickness of 100 μm. The detector was located at a distance of 24 cm from the target center and mounted on a rotatable arm inside the chamber, thus enabling it to be moved in the angular range of 10° to 75° in the laboratory system. More information about the experimental setup and the scattering chamber used in the experiment can be found in earlier work [12, 13]. The instruments used for processing the detector signals, corresponding to the reaction products, further included electronic components from ORTEC and CANBERRA with MAESTRO [11] software for recording and processing of the spectra of the nuclear processes. The angular distribution for $^{10}$B ($^{16}$O,$^{16}$O)$^{10}$B system was measured in the angular range of ~ 35°–120° in the center of mass system. Beam current was measured using a Faraday Cup to be nearly 45 nA during the experiment. Energy spectra of scattered particles were measured using a silicon surface barrier detector (ORTEC) with a sensitive layer thickness of 100 μm. The energy resolution of the registration system was 250–300 keV, which is mainly determined by the energy spread of the primary beam. The detector was located at a distance of 24 cm from the scattering region and had the opportunity to move in the angular range from 10° to 75° in the laboratory system.

The $^{16}$O beam passed through three collimators of 1.5 mm diameter and was focused on the target to a spot diameter of ≈ 3.9 mm. Figure 2 shows the spectrum for the $^{10}$B ($^{16}$O,$^{16}$O)$^{10}$B elastic scattering at a detector angle of 24°. Final normalization of the absolute cross-sections was done by comparing the measurements at the most forward angles, where Rutherford scattering dominates, with the optical model predictions which, in this angular region, depend only weakly on the potential parameters. More information about the experimental setup and the scattering chamber applied in the experiment could be found in Ref. [12,13]. We estimated the systematic error of measured cross-sections to be no larger than 10%. The statistical error was 1–5% during our measurements in the region of the forward hemisphere and increased at backward angles but nowhere exceeded 10%. The error bars on the cross-sections are smaller than the size of the experimental points.

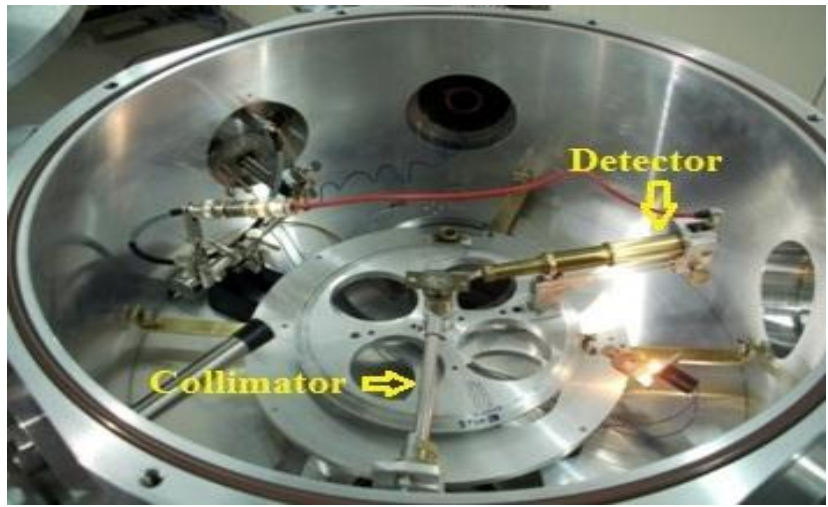

**Fig. 1:** The scattering chamber applied in our experiment performed at cyclotron DC−60

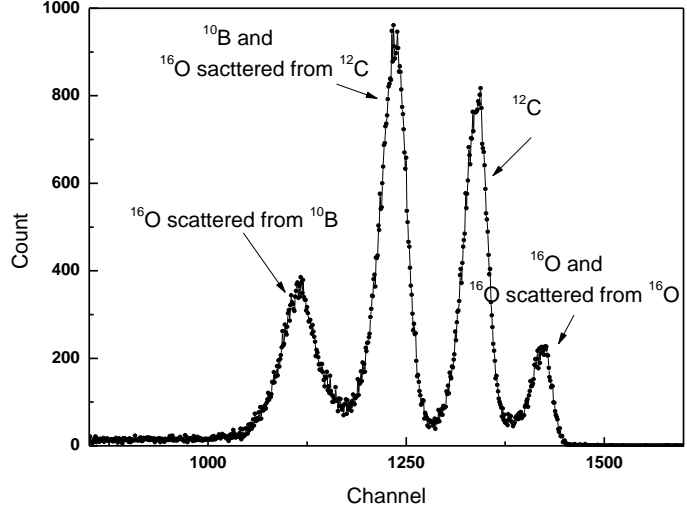

**Fig. 2:** Spectrum for $^{10}$B ($^{16}$O,$^{16}$O)$^{10}$B elastic scattering at angle 24° and at energy 24 MeV

### III. Theoretical Analysis

The folding model is well known as a powerful tool for analyzing the nucleus-nucleus scattering at low and intermediate energies. It directly links the nuclear density profile with the scattering cross-sections and is therefore quite appropriate for carrying out a theoretical study of the experimental data for the $^{16}$O+$^{10}$B system at $E_{\text{lab}}(^{16}\text{O}) = 24$ MeV, from a microscopic point of view, via a double-folding (DF) model. The real part of the potential is constructed by folding the nucleon-nucleon interaction into the nucleon densities of the projectile $\rho_p(r_1)$ and target nuclei $\rho_t(r_2)$ in their ground states using the code DFMSPH [14]. The resultant potential is then multiplied by a renormalization factor which fits the experimental elastic scattering cross section. The real part of the nucleus-nucleus potential in the DF model is written as [15–17]

$$V_{DF}(R) = \int \rho_P(r_1)\rho_T(r_2)\upsilon_{nn}(s)d\vec{r_1}d\vec{r_2} \quad \text{MeV} \tag{1}$$

where $\upsilon_{nn}(s)$ is the effective nucleon-nucleon interaction potential and $s = |\vec{R} - \vec{r}_1 + \vec{r}_2|$ is the distance between the two nucleons. $\upsilon_{nn}(s)$ was taken to be of the DDM3Y1 form, $v_{EX}(s)$, based on the M3Y-Paris potential, $v_D(s)$:

$$v_D(s) = 11061.625 \frac{\exp(-4s)}{4s} - 2537.5 \frac{\exp(-2.5s)}{2.5s}, \text{MeV}$$

$$v_{EX}(s) = -1524.25 \frac{\exp(-4s)}{4s} - 518.75 \frac{\exp(-2.5s)}{2.5s} - 7.8474 \frac{\exp(-0.7072s)}{0.7072s}, \text{MeV} \quad (2)$$

The M3Y-Paris potential is scaled by an explicit density-dependent function $F(\rho)$:

$$v_{D(EX)}(\rho, s) = F(\rho) v_{D(EX)}(s), \quad (3)$$

where $v_{D(EX)}$ are the direct and exchange components of the M3Y-Paris, $\rho$ is the nuclear matter density and $s$ is the distance between the two interacting nucleons. The density-dependent function $F(\rho)$ was taken to have an exponential dependence as follows:

$$F(\rho) = C[1 + \alpha \exp(-\beta \rho) - \gamma], \quad (4)$$

The parameters $C$, $\alpha$, $\beta$, and $\gamma$ of the DDM3Y1 potential, listed in Table I, were taken from earlier work [18]. These parameters give the corresponding value of the nuclear incompressibility, $K$, in the Hartree–Fock (HF) calculation of nuclear matter [19].

**Table I**: Parameters of density-dependence function $F(\rho)$

| Interaction Model | c | α | β (fm$^3$) | γ (fm$^{3n}$) | K (MeV) |
|---|---|---|---|---|---|
| **DDM3Y1** | 0.2963 | 3.7231 | 3.7384 | 0.0 | 176 |

The density distribution of $^{16}$O is expressed using a modified form of the Gaussian shape as $\rho(r) = \rho_0 (1 + wr^2) \exp(-\beta r^2)$, where $\rho_0 = 0.1317$, $w = 0.6457$, and $\beta = 0.3228$ [20]. The density distribution of $^{10}$B is calculated using a modified form of the harmonic oscillator function $\rho(r) = \rho_0 \left[1 + \alpha \left(\frac{r}{a}\right)^2\right] \exp\left(-\left(\frac{r}{a}\right)^2\right)$, where $\rho_0 = 0.1592$, $a = 1.71$, and $\alpha = 0.837$ [21]. The theoretical analysis, using the experimental data for constraining the model parameters, was done within the framework of the double-folding optical potential (DFOP) model. In this model, the real part of the potential was derived on the basis of the double-folding model, as discussed above, and the imaginary part was taken to have the standard Woods-Saxon form. Thus, the total interaction potential in this case has the following shape:

$$U(R) = V_C(R) + Nr\, V^{DF}(R) - iW(R) \tag{5}$$

$V_C(R)$ is the Coulomb potential of a uniform charged sphere. The calculated potentials for the un-normalized real part at energies of $E_{lab}$=24, 36.58, 48.49, and 64 MeV are shown in Fig. 3. It is evident that all potentials are close to each other in the surface (r > 5 $fm$) as well as the inner regions.

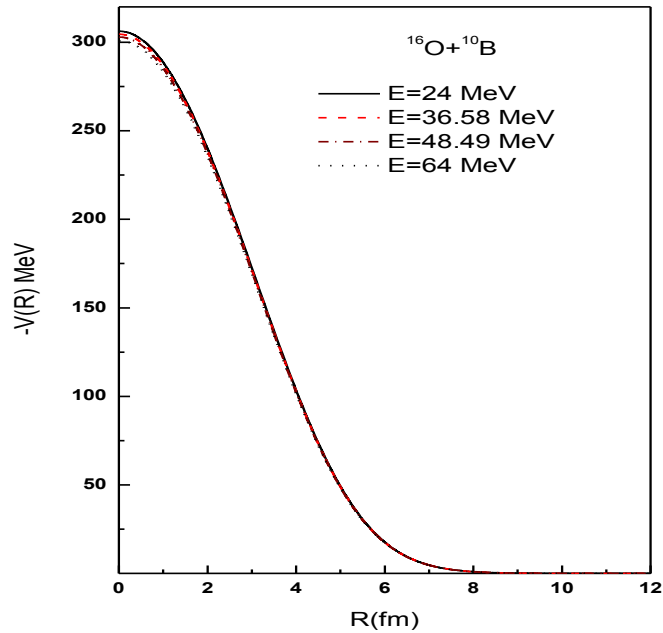

**Fig. 3:** The calculated potential for the real part at $E_{lab}$=24, 36.58, 48.49, and 64.0 MeV using double folding based on DDM3Y1 interaction.

## IV. Results and Discussion

### A. Elastic scattering $^{10}B(^{16}O,^{16}O)^{10}B$

In addition to our experimental measurements for the $^{16}O+^{10}B$ system at $E_{lab}$=24 MeV, we analyzed the aforementioned nuclear system at different energies. A comparison between the experimental angular distributions at energies 21.37, 23.27, 26.0, 27.3 [7] and 24.0 MeV with the corresponding theoretically calculated distributions, obtained using DFOP model, is shown in Fig. 4. The experimental data in this energy range do not show any increase in the cross-section at backward angles which simply implies that the transfer phenomenon could not be observed at these energies that are very close -slightly above and below- the Coulomb barrier energy ($V_{CB}$) of 25.75 MeV for the $^{16}O+^{10}B$ system. The optimal potential parameters used in calculations are listed in Table II along with the values of the real volume integral ($J_V$) and imaginary volume integral ($J_W$).

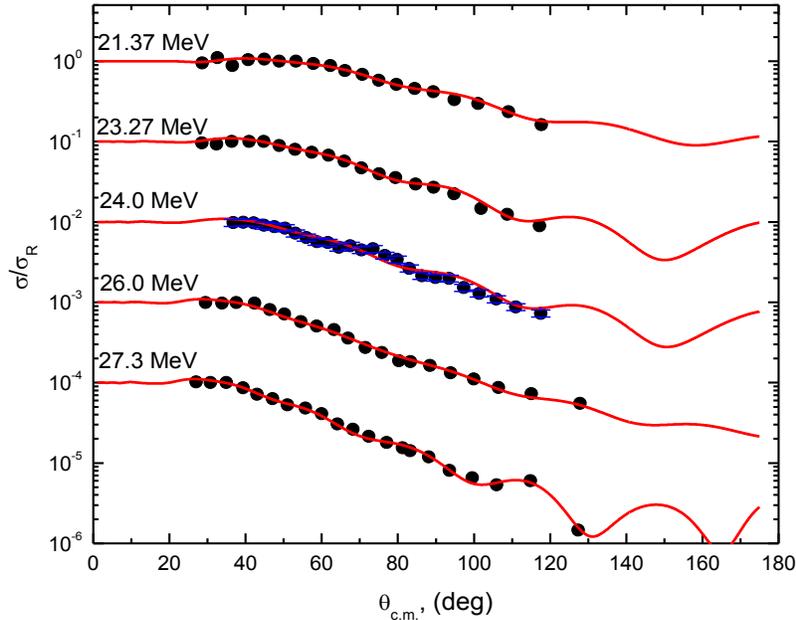

**Fig. 4:** Comparison between experimental angular distributions data (solid black circles) for $^{10}B$ $(^{16}O,^{16}O)^{10}B$ elastic scattering and the theoretical calculations (solid red curves) using DFOP model at $E_{lab}$= 21.37, 23.27, 24.0, 26.0 and 27.3 MeV. Note that datasets at different energies have been displaced by successive factors of $10^{-1}$ for the sake of clarity.

The best fit to the experimental data was obtained by minimizing the $\chi^2/N$ (where $N$ stands for the number of data points). The experimental data were fitted using four parameters: the renormalization factor ($N_R$) for the real part of the potential, derived on the basis of the double-folding model, and the depth ($W$), radius ($r_w$) and diffuseness ($a_w$) for the imaginary part of potential. The parameters $r_w$ and $a_w$ were kept constant during the search allowing only two parameters $N_R$ and $W$ to be changed until the least $\chi^2/N$ value was achieved. Although we sacrificed the quality of fitting at some energies by using this technique, we still could obtain reliable energy dependence of $N_R$ and $W$.

The total reaction cross sections ($\sigma_R$), obtained from the calculations and listed in Table II, are plotted as a function of energy $E$, as shown in Fig. 5. The systematic variation of $\sigma_R$ with $E$ is typically the same as that reported previously by Anjos *et al.* [9]. Our extracted values of $\sigma_R$ are 1.2–2.5 times higher than the values reported earlier [9] and unfortunately, no other reported values of $\sigma_R$ from the previous studies seem to match with our results. Quadratic fit to $\sigma_R$ was obtained using

$$\sigma_R = a + bE + cE^2 \tag{6}$$

where $a = -759.4$ ($-955.3$), $b = 72.7$ (63.3) and $c = -0.6$ ($-0.5$) for the data obtained in the present work and those obtained from earlier work [9], respectively. We note that the value of $\sigma_R$ increases with increasing energy at low energies (< 60 MeV) and almost saturates at energies higher than 60 MeV.

**Table II:** Optimal potential parameters for $^{16}$O+$^{10}$B nuclear system at different energies, together with *SA* values extracted from the DWBA analysis. Note that $\chi^2/N$ values refer to $\theta_{c.m.}<90°$ for the elastic scattering and the full angular range for the DWBA calculations. Coulomb radius parameter was fixed at 1.25 fm.

| E | | $N_R$ | W (MeV) | $r_w$ (fm) | $a_w$ (fm) | $\chi^2/N$ | SA | $\sigma_R$ (mb) | $J_V$ (MeV.fm$^3$) | $J_W$ (MeV.fm$^3$) |
|---|---|---|---|---|---|---|---|---|---|---|
| 21.37 | Elastic | 0.969 | 6.89 | 1.35 | 0.466 | 0.59 | | 469.8 | 547.5 | 47.76 |
| 23.27 | Elastic | 0.969 | 6.89 | 1.35 | 0.466 | 0.81 | | 595.8 | 547.5 | 47.76 |
| 24.0 | Elastic | 0.959 | 7.48 | 1.35 | 0.466 | 1.65 | | 637.3 | 539.9 | 51.85 |
| 26.0 | Elastic | 0.957 | 10.75 | 1.35 | 0.466 | 0.46 | | 744.5 | 539.7 | 74.52 |
| 27.3 | Elastic | 1.0 | 8.15 | 1.35 | 0.466 | 0.42 | | 810.8 | 564 | 56.49 |
| 36.58 | Elastic | 0.925 | 11.19 | 1.35 | 0.466 | 0.49 | | 1089 | 518.9 | 77.57 |
| | DWBA | | | | | 8.34 | 1.22 | | | |
| 41.99 | Elastic | 0.986 | 11.19 | 1.35 | 0.466 | 4.17 | | 1203 | 552.2 | 77.57 |
| | DWBA | | | | | 11.67 | 1.34 | | | |
| 48.49 | Elastic | 0.999 | 13.73 | 1.35 | 0.466 | 7.1 | | 1302 | 557.4 | 95.18 |
| | DWBA | | | | | 17.61 | 1.36 | | | |
| 64.0 | Elastic | 0.93 | 11.63 | 1.35 | 0.466 | 17.4 | | 1399 | 496.0 | 80.62 |
| | DWBA | | | | | 24.85 | 1.44 | | | |

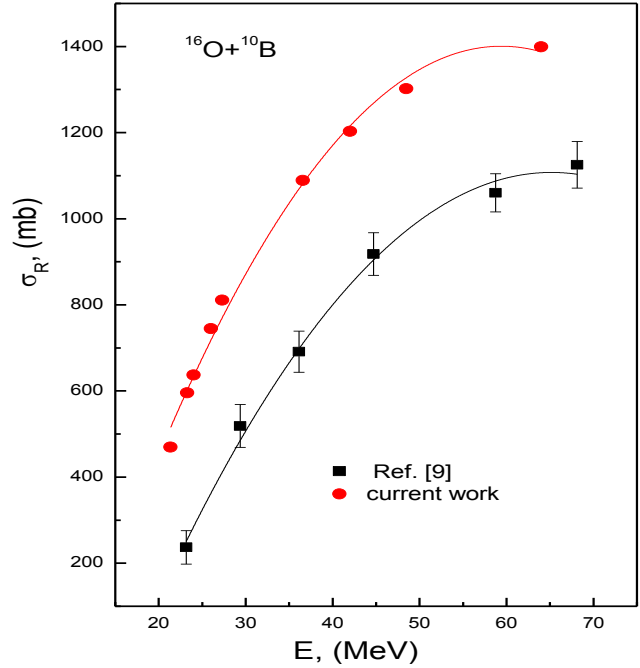

**Fig. 5:** Energy dependence of the extracted total reaction cross sections for the $^{10}$B ($^{16}$O,$^{16}$O)$^{10}$B elastic scattering from the current work and those from an earlier report [9]. The lines are the fit results.

The experimental data at higher energies of 36.58, 41.99, 48.49, and 64.0 MeV [8, 9] show a significant increase in the cross-sections at backward angles. Such an increase could be investigated in terms of the $^6$Li cluster transfer between $^{16}$O and $^{10}$B. To this end, firstly, the experimental angular distributions at the aforementioned energies were analyzed up to angles < 90$^o$ so as to exclude the effect of cluster transfer which causes a significant growth in cross-sections at backward angles. Data at forward angles corresponding to pure elastic scattering were analyzed using the DFOP model employing the FRESCO code [22]. The potential parameters extracted from the analysis are presented in Table II.

### B. *Elastic transfer $^{10}$B($^{16}$O, $^{10}$B)$^{16}$O*

As mentioned earlier, the gross features in cross-sections at backward angles can be explained to be due to cluster transfer. DWBA calculations were performed to explore the possibility of $^{16}$O to be treated as ($^{10}$B–Core) + ($^6$Li–valence). In this case, the exchange of a $^6$Li cluster between the two interacting nuclei leads to an exit channel that is physically indistinguishable from the entrance channel. Thus, the differential cross sections will be the square of the sum of amplitudes from the pure elastic scattering and the exchange mechanism of the cluster transfer as follows, $\frac{d\sigma_{el}}{d\Omega} = |f_{el}(\theta) + e^{i\alpha}Sf_{DWBA}(\pi - \theta)|^2$, where $f_{el}(\theta)$ is the elastic scattering amplitude, $f_{DWBA}(\pi - \theta)$ is the amplitude calculated in the distorted wave method with the replacement $\theta \rightarrow \pi - \theta$, $S$ is the product of the two spectroscopic amplitudes (*SA*) of the transferred particle in the initial and final states which are the same as in the case of elastic transfer.

Calculations of transfer were performed using the same optimal potential parameters obtained by fitting the experimental data in the forward hemisphere up to 90°. The bound state wave function for the relative motion of $^6$Li and $^{10}$B in the cluster plus core configuration in $^{16}$O was defined by a Woods-Saxon potential with a fixed radius $R = 1.25\ (A_p^{1/3} + A_t^{1/3})$ fm and diffuseness $a = 0.65$ fm. The potential depth was adjusted to reproduce the binding energy of 30.874 MeV of the cluster. The number of nodes (*N*) were determined using the Talmi–Moshinsky formula [23], $2(N-1) + L = \sum_{i=1}^{n} 2(n_i - 1) + l_i$, where $n_i, l_i$ are quantum numbers of the nucleons in the cluster and *L* is orbital angular momentum of the cluster. Cluster quantum numbers for the overlaps $\langle ^{16}\text{O} | ^{10}\text{B} + ^6\text{Li} \rangle$ used in our calculations are listed in Table III. The same potential parameters were taken for the entrance channel ($^{16}$O+$^{10}$B) and the exit channel ($^{10}$B+$^{16}$O). A comparison between the angular distributions at energies 36.58, 41.99, 48.49 and 64 MeV and theoretical calculations using the DFOP model is shown in Fig. 6 for both $^{10}$B ($^{16}$O,$^{16}$O) $^{10}$B pure elastic scattering (angles< 90°) as well as for the elastic transfer $^{10}$B ($^{16}$O,$^{10}$B) $^{16}$O after including the effect of $^6$Li cluster transfer.

The spectroscopic amplitude was taken as a free parameter that was varied in order to give the best agreement between the theoretical calculations and the experimental data and consequently the least $\chi^2/N$ value. The variation of the extracted *SA* with $\chi^2/N$ at $E_{lab}$= 36.58, 41.99, 48.49 and 64.0 MeV is shown in Fig. 7. The extracted spectroscopic amplitude for the configuration $^{16}$O→$^{10}$B+$^6$Li is 1.34 ±0.091 and its values at the different energies are listed in Table II.

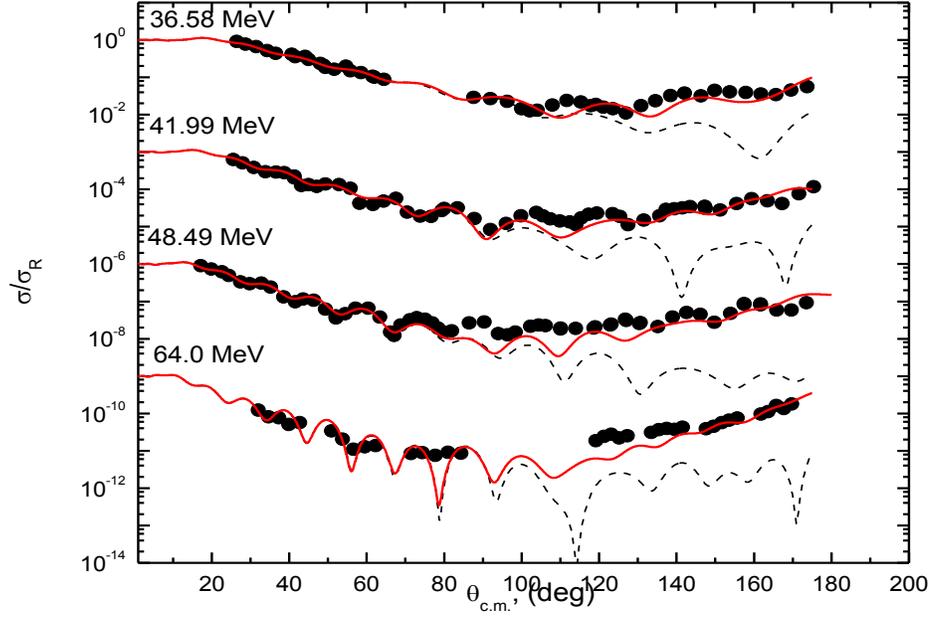

**Fig. 6:** The comparison between the experimental data (solid black circles) and calculations for the $^{10}$B ($^{16}$O,$^{16}$O)$^{10}$B elastic scattering at $E_{lab}$= 36.58, 41.99, 48.49, and 64.0 MeV. The dashed black curves denote that pure optical model fits the data for angles $\theta_{c.m.}$< 90°. The solid red curves denote that the results of DWBA calculations including the $^{10}$B ($^{16}$O,$^{10}$B)$^{16}$O elastic transfer process. It should be noted that datasets at different energies have been displaced by successive factors of $10^{-3}$ for the sake of clarity.

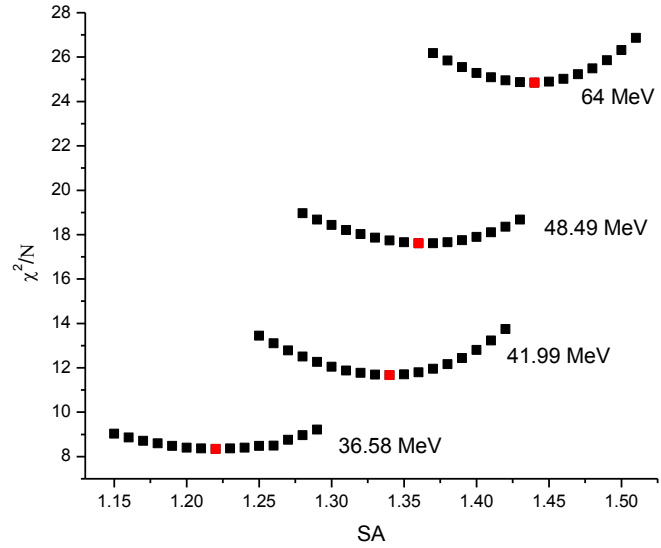

**Fig. 7:** Variation of $\chi^2/N$ with extracted *SA* at $E_{lab}$= 36.58, 41.99, 48.49, and 64.0 MeV

**Table III:** *N, L, S,* and *J* for the overlaps used in our calculations

| Overlap | N<br>Number of nodes | L | S | J = L+S | B.E.<br>MeV |
|---|---|---|---|---|---|
| $\langle {}^{16}O | {}^{10}B+{}^{6}Li \rangle$ | 3 | 2 | 1 | 3 | 30.874 |

To check the reliability of the real and imaginary parts of the potential, we have applied the dispersion relation to the values of their volume integrals. The volume integrals of the real and imaginary potentials and the dispersion relation [24, 25] between them have been calculated by using the following formulae:

$$J_{V,W}(E) = \frac{4\pi}{A_P A_T} \int_0^R V,W(r,E) r^2 dr, \tag{7}$$

$$V_N(E) = V_R + \Delta V(E) = V_R - (W/\pi)\left[\varepsilon_a \ln|\varepsilon_a| - \varepsilon_b \ln|\varepsilon_b|\right] \tag{8}$$

Here $\varepsilon_i = (E - E_i)/(E_b - E_a)$ with $i = a, b$, respectively. The energy $E_a$ is assumed to be the value at which the imaginary potential vanishes and $E_b$ is the reference energy. The parameter values of $E_a = 10$ MeV, $E_b = 100$ MeV, $V_R = 564$ MeV, and $W = 34$ MeV were used for our calculated potentials and the volume integrals thus obtained are plotted in Fig. 8. From the figure, it can be seen that the real volume integral has apparent energy dependence, where $J_V$ decreases as the energy increases while the imaginary volume integral strength, $J_W$, increases quickly in accordance with the dispersion relation curve.

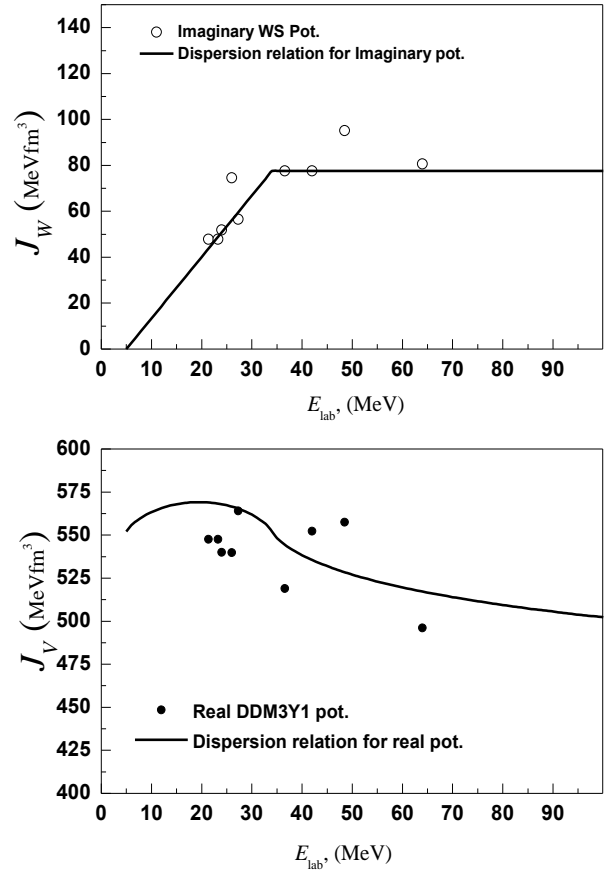

**Fig. 8:** Volume integrals for the potentials obtained for the $^{10}$B ($^{16}$O,$^{16}$O)$^{10}$B. Full and empty circles are the results from the microscopic calculations in comparison with the dispersion relation between real and imaginary components of the nuclear potential.

## V. Summary

We measured the angular distribution for $^{16}$O elastically scattered from $^{10}$B at $E_{lab}$= 24 MeV. The measured cross-sections are found to decrease steadily with increasing scattering angle. The same behavior has also been observed at energies of $E_{lab}$= 21.37, 23.27, 26.0 and 27.3 MeV. Data at higher energies of $E_{lab}$=36.58, 41.99, 48.49 and 64.0 MeV show a significant increase in cross-sections at backward angles. This observation was previously interpreted in terms of the compound elastic process using the statistical model. In the present work, we examined the effect of $^6$Li exchange between $^{16}$O and $^{10}$B and its effect on the cross-sections at backward angles. We have also extracted the values of the reaction cross-section, $\sigma_R$, and compared them with the corresponding values obtained from previous measurements as well as with the dispersion relation curve. Additionally, the *SA* for the configuration $^{16}$O→$^{10}$B+$^6$Li is extracted to be 1.34 ±0.091. Furthermore, the cluster structure of $^{16}$O as a core ($^{10}$B) plus a valence particle ($^6$Li) orbiting the core is observed to successfully reproduce the significant rise in cross-sections at backward angles.

**Acknowledgements** This work has been support in part by the Ministry of Education and Science of the Republic of Kazakhstan (grant No. AP05132062). A. A. Ibraheem extend his appreciation to the Deanship of Scientific Research at King Khalid University for funding this work through research groups program under grant number R.G.P.1/118/40